\documentclass{svmult}
\usepackage{makeidx}         
\usepackage{graphicx}        
\usepackage{multicol}        
\usepackage[bottom]{footmisc}

\makeindex             

\usepackage{epsfig}
\usepackage{graphics}
\usepackage{dcolumn}
\usepackage{bm}

\def\cm-1{cm$^{-1}$}

\def\scbo{SrCu$_{2}$(BO$_{3}$)$_{2}$\,}

\begin{document}

\title*{Collective Magnetic Excitations in \scbo}
\author{A. Gozar\inst{1,2} \and G. Blumberg\inst{1,*}}
\institute{
Bell Laboratories, Lucent Technologies, Murray Hill, NJ 07974, USA \and 
University of Illinois at Urbana-Champaign, Urbana, IL 61801, USA
\texttt{
\begin{center}
    (\textit{Frontiers in Magnetic Materials} (Ed. A.V. Narlikar), 
    Spinger-Verlag 2005, pp. 735-754).
\end{center}
}} 
%
%
\maketitle

\section{Introduction: Why \scbo ?}

Several properties make \scbo a unique and one of the most interesting quantum magnets~\cite{MiyaharaJPCM03}.
This compound is a 2D spin system  with a disordered ground state even at very low temperatures and a spin gap of about 24~\cm-1 (3~meV) in the magnetic excitation spectrum~\cite{MiyaharaJPCM03,KageyamaPRL99}.
It has been established that the elementary excitations in this system, which consist of a transition from the singlet (S = 0) ground state to the lowest energy triplet (S = 1) state, are local, weakly dispersive in the reciprocal space while many-'particle' magnetic states are more mobile, i.e. show more dispersion.
The strengths of the relevant magnetic interactions place this compound close to a quantum critical point (QCP) separating the gapped phase from a gapless state having long range magnetic AF order.
Moreover data in high magnetic fields show plateaus at commensurate (1/8, 1/4 and 1/3) values of the saturation magnetization~\cite{KageyamaPRL99,OnizukaJPSJ00,KodamaScience02}.
The plateau states can be thought of as crystalline arrangements of magnetic moments separating regions of continuous rise in magnetization, these latter regions allowing for an interpretation in terms of Bose-Einstein condensation of triplet excitations~\cite{RiceScience02}.
It has also been suggested~\cite{SriramPTP02} that doping in this system (regarded as a Mott-Hubbard insulator) may lead to a superconducting phase mediated by antiferromagnetic (AF) fluctuations, a mechanism similar in spirit to one of the scenarios proposed for the high T$_{c}$ cuprates~\cite{AndersonScience87}.

The first synthesis of \scbo was achieved in 1991 by Smith and collaborators~\cite{SmithJSSC91} but the authors did not elaborate on the magnetic properties. This compound was rediscovered in 1999 by Kageyama \emph{et al.}~\cite{KageyamaPRL99}, who also pointed out the outstanding magnetic properties and the importance of \scbo in the physics of low dimensional quantum magnets.
The crystal belongs to the tetragonal symmetry and it has a layered structure in which Cu(BO$_{3}$) units are separated by closed shell Sr$^{2+}$ atoms, see Fig.~1.
At T~=~395~K the system undergoes a 2$^{nd}$ order phase transition from the space group $I4/mcm$ to $I\bar{4}2m$ on cooling from high temperature.
In the $I4/mcm$ phase the planes containing the Cu atoms are flat and they form mirror symmetry elements.
Below 395~K the transition can be intuitively understood as the buckling of the Cu planes which lose their mirror symmetry.

The magnetic properties of \scbo are determined by the $S = 1/2$ spins sitting on the Cu$^{2+}$ sites.
The crystal structure imposes a distribution of the magnetic moments in weakly coupled 2D layers which define the $(ab)$ planes.
In each of these sheets the spins form orthogonal dimer lattices and the $c$-axis is perpendicular to them.
The in-plane magnetic interactions can be described in an effective spin model by taking into account the magnetic exchange between nearest neighbor Cu spins bridged O atoms forming a 147$^{\circ}$ Cu-O-Cu bond which points toward an AF coupling denoted by $J_{1}$, see Fig~\ref{f12}.
Besides this interaction one can consider the next nearest neighbor super-exchange taking place $via$ the BO$_{3}$ complexes.
Curie-Weiss type fits of the high temperature magnetization data suggest that this coupling, denoted by $J_{2}$, is also AF.
Indeed, as we will discuss later, these two terms seem to capture many aspects regarding the magnetic properties of \scbo.
The addition of a weak inter layer exchange $J'$ to the above two terms constitute the starting point for treating the 3D spin dynamics in this compound.
\begin{figure}[h]
\centerline{
\epsfig{figure=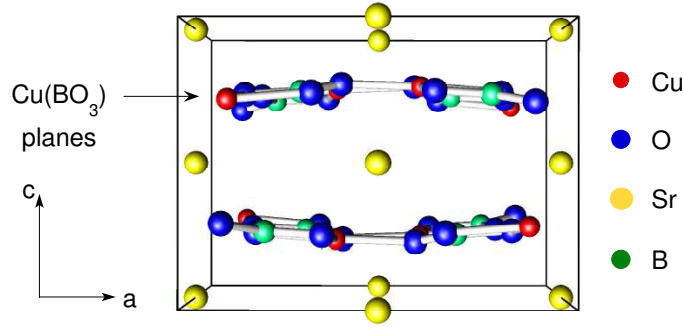,width=90mm}
}
\caption{
3D view of the \scbo crystal showing the layered structure along the $c$-axis. Above 395~K the Cu(BO$_{3}$) planes are flat, mirror symmetry elements.
}
\label{f11}
\end{figure}

The schematic of the magnetic lattice is shown in Fig.~2 where the nearest neighbor AF bonds are represented by a solid line while the next nearest neighbor interaction is shown by dashed lines.
Note an important property of this lattice: ignoring the solid bonds (the $J_{1}$ interactions), the topology of the spin structure, determined by the $J_{2}$ bonds, is equivalent to that of a 2D square lattice.
The corresponding Hamiltomian for one Cu(BO$_{3}$) plane reads:
\begin{equation}
\hat{H} = J_{1} \sum_{(i,j) \ NN} {\bf S}_{i} \cdot {\bf S}_{j} + J_{2} \sum_{[i,j] \ NNN} {\bf S}_{i} \cdot {\bf S}_{j} 
\label{e11}
\end{equation}
Here $i$ and $j$ are nearest neighbor (NN) or next nearest neighbor (NNN) Cu sites.
\begin{figure}[b]
\centerline{
\epsfig{figure=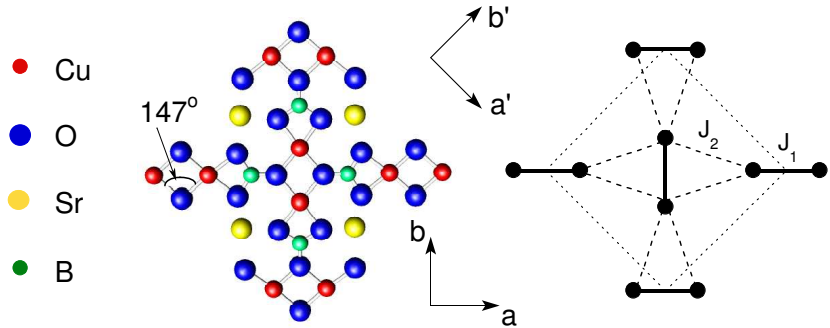,width=110mm}
}
\caption{
Left: The $(ab)$ planes and crystal axes notation.
Right: The effective magnetic lattice with the unit cell of the dimer structure being shown by the short dashed square.
The circles are the $S = 1/2$ Cu spins, the solid and long dashed lines represent the intra-dimer ($J_{1}$) and inter-dimer ($J_{2}$) AF superexchange interactions.
Below 395~K~\cite{SpartaEPJB01} the vertical and horizontal dimers become slightly non-coplanar, see Fig.~\ref{ f11}.
}
\label{f12}
\end{figure}

Why is a Raman study of this compound interesting?
Several magnetic properties, features of the excitation spectrum or the occurrences of the magnetization plateaus, can be explained by taking into account the $J_{1}$ and $J_{2}$ interactions.
As will be discussed in the review of the main theoretical results, this compound is believed to be characterized by a ratio $x = J_{2}/J_{1}$ close to 0.7, which is the value around which the system ground state changes.
One reason to study this compound is that an accurate determination of this ratio is still missing and this is important due to the proximity to the QCP.
In this regard the recent observation of a mode below the spin gap brings into question the quantitative determination of the AF exchange parameters of the system since the existence of such excitation has not been predicted by theory.
Another reason is brought about by a different set of questions related to the way the external radiation field couples to magnetic excitations.
Magnetic modes in the triplet sector have been probed spectroscopically, besides inelastic neutron scattering (INS)~\cite{KageyamaPRL99,CepasPRL01}, also by electron spin resonance (ESR)~\cite{NojiriJPSJ99,NojiriJPSJ03} and infra-red (IR)~\cite{ToomasPRB00,ToomasPrivate} absorption experiments.
The nature of the mixing interactions originating in spin-orbit coupling which usually allows transitions from the singlet ($S = 0$) to excited ($S = 1$) states is still to be understood.
Our approach, which is a study of collective magnetic excitations in terms of symmetry, resonance and coupling mechanisms in external magnetic fields, is illuminating in this respect.

This chapter will be focussed on the low temperature properties of \scbo.
The discussion of the low temperature phononic excitations as seen in Raman spectra in the next part will be followed by a review of experimentally found magnetic properties with an emphasis on spectroscopic techniques.
Then we will present a description of the basic properties emerging from the Hamiltonian~(\ref{e11}) which describes a Shastry-Sutherland lattice and the structure of the magnetic excitation spectrum.
The effects of other interactions, in particular of the antisymmetric Dzyaloshinskii-Moriya (DM) terms, will be described.
Then we will present our low temperature Raman data and discuss our results in reference to the open questions mentioned above.

\section{Low Temperature Phononic Spectra in \scbo}

The analysis of the low temperature lattice dynamics is important for several reasons.
In general this is because there are many examples of low dimensional crystalline compounds which undergo transitions to phases in which the appearance of a spin gap in the magnetic excitation spectrum is accompanied by real space lattice symmetry breaking due to spin-phonon coupling.
The lowering of the crystal symmetry may involve newly allowed phononic modes which could be checked directly in the Raman spectra.
In the particular case of \scbo spin-lattice interaction has been suggested to be relevant to the magnetic dynamics at low temperatures and/or high magnetic fields.
In order to explain the selection rules of the transitions seen in IR absorption, spin-phonon induced antisymmetric DM spin interactions have been invoked~\cite{ToomasPrivate}.
It has been argued that this coupling will induce virtual phonon transitions which will instantaneously lower the crystal symmetry, allowing for non fully-symmetric effective singlet triplet mixing terms.  
On the other hand the spin-lattice coupling was taken into account in order to be able to describe the spin density profile at high fields in the magnetization plateaus states.
NMR data brought evidence for the broken translational symmetry at the 1/8 plateau by the existence of at least 11 nonequivalent Cu sites~\cite{KodamaScience02}. 
The role of the phonons coupled adiabatically to the spin degrees of freedom in this case is to lift the degeneracy of the ground state, picking a state with a certain magnetization texture and allowing the ground state magnetization to be a defined quantity.

\begin{figure}[b]
\centerline{
\epsfig{figure=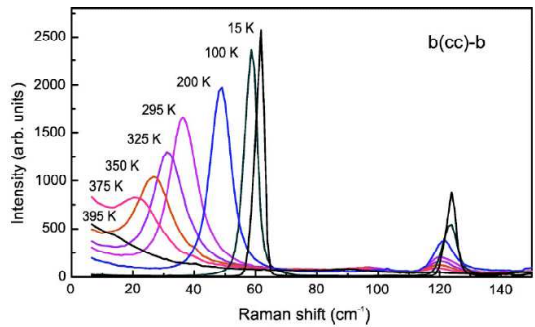,width=70mm}
}
\caption{
The fully symmetric soft mode below the $I4/mcm$ - $I\bar{4}2m$ transition from Ref.~\cite{ChoiPRB03} seen in $cc$ polarization.
Note the strong two-phonon excitation around 120~\cm-1.
}
\label{f13}
\end{figure}

So far the study of phononic excitations has been especially focussed in relation to the structural phase transition at T~=~395~K \cite{SpartaEPJB01,ChoiPRB03}.
The lattice soft mode of the 2$^{nd}$ order transition from the $I4/mcm$ to the $I\bar{4}2m$ group can be seen in Fig~\ref{f13}.
The phonon which condenses belongs to the B$_{1u}$ representation of the higher symmetry group and, in terms of Cu atoms, it involves mainly an alternate displacement along the $c$-axis of the nearest neighbor dimers, see Fig~\ref{f11}.

A symmetry analysis of the phononic excitations in \scbo is done in the following.
The unit cell contains 4 formula units and a total of 44 atoms.
The number of $k = 0$ modes is given by $3 \times 44 / 2 = 66$, the factor of 2 coming from the fact that the unit cell is body centered.
\begin{table}[t]
\caption[]
{
Symmetry of the phononic excitations of \scbo in the high ($I4/mcm$, point group $D_{4h}$) and low ($I\bar{4}2m$, point group $D_{2d}$) temperature phases.
The first column contains the atom types and the rest of the columns correspond to irreducible representations of the point groups. 
}
\vspace{0.2cm}
\begin{center}
\begin{tabular}{c|c|c|c|c|c|c|c|c|c|c||c|c|c|c|c}
 & \multicolumn{10}{c}{$I4/mcm$} & \multicolumn{5}{c}{$I\bar{4}2m$} \\
Atom & A$_{1g}$ & A$_{2g}$ & B$_{1g}$ & B$_{2g}$ & E$_{g}$ & A$_{1u}$ & A$_{2u}$ & B$_{1u}$ & B$_{2u}$ & E$_{u}$ & A$_{1}$ & A$_{2}$ & B$_{1}$ & B$_{2}$ & E \\
\hline
\hline
Cu & 1 & 1 & 1 & 1 & 1 & - & 1 & 1 & - & 2 & 2 & 1 & 1 & 2 & 3 \\ 
\hline
O1 & 1 & 1 & 1 & 1 & 1 & - & 1 & 1 & - & 2 & 2 & 1 & 1 & 2 & 3\\
\hline
O2 & 2 & 2 & 2 & 2 & 2 & 1 & 1 & 1 & 1 & 4 & 3 & 3 & 2 & 2 & 6\\
\hline
B & 1 & 1 & 1 & 1 & 1 & - & 1 & 1 & - & 2 & 2 & 1 & 1 & 2 & 3\\
\hline
Sr & - & 1 & - & - & 1 & - & 1 & - & - & 1 & - & 1 & - & 2 & 2\\
\end{tabular}
\end{center}
\label{t11}
\end{table}
In this case, the symmetry lowering at 395~K does not change the number of atoms in the unit cell.
The associated point group of the $I4/mcm$ space group is $D_{4h}$ while $D_{2d}$ corresponds to $I\bar{4}2m$.
The analysis is based on the tables in Ref.~\cite{PortoBauman} and uses the site symmetry approach.
Oxygen atoms mediating the intra-dimer superexchange (O1) and the Oxygens allowing for the inter-dimer superexchange (O2) occupy a different symmetry positions.
In the high temperature phase the Cu, O1 and B atoms have $C'_{2v}$, O2 atoms have $C_{s}$ and Sr atoms occupy $D_{4}$ site symmetries respectively.
Table~\ref{t11} summarizes the number of modes corresponding to each atom both above and below the transition.
The part of the table related to the $I\bar{4}2m$ group can be easily inferred from the analysis of the high temperature phase.
One has to drop the 'u' and 'g' indices (corresponding to odd and even modes) because the inversion symmetry is lost and by using compatibility tables which show that the representations ${A_{1u}}, A_{2u}, B_{1u}, B_{2u}$ of the $D_{4h}$ point group become ${B_{1}}, B_{2}, A_{1}, A_{2}$ representations (in this order) in $D_{2d}$.
Both the $E_{u}$ and $E_{g}$ representations remain double degenerate.

\begin{figure}[b]
\centerline{
\epsfig{figure=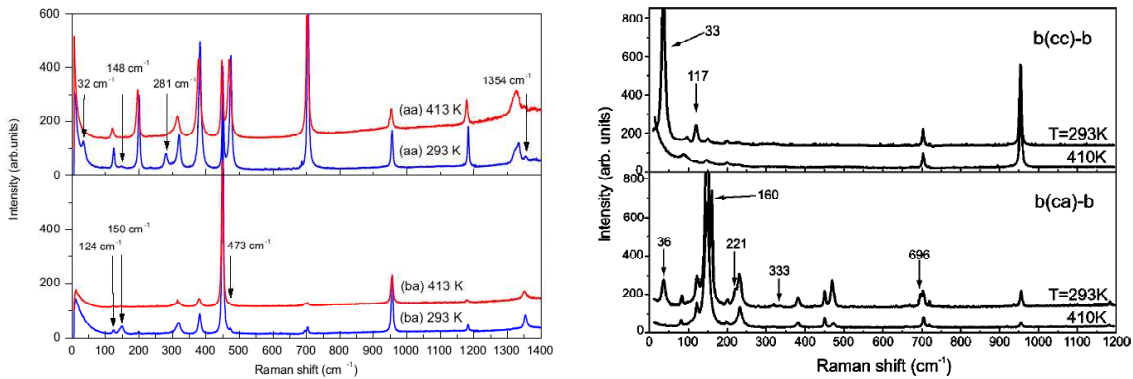,width=150mm}
}
\caption{
Phononic data from Ref.~\cite{SpartaEPJB01} (left) and Ref.~\cite{ChoiPRB03} (right) showing in plane and out of plane polarized spectra above and below the structural transition.
Newly allowed Raman phonons are marked by arrows. 
}
\label{f14}
\end{figure}

Inversion symmetry breaking results in the fact that certain phonons, which were dipole active, become Raman allowed.
The appearance of several new modes below 395~K is shown in Fig.~\ref{f14} where data from Refs.~\cite{SpartaEPJB01,ChoiPRB03} is reproduced.
Notable are the shoulders appearing around 160 and 220~\cm-1 below 395~K in $ca$ polarization (which probes double degenerate modes with E symmetries) suggesting either an almost degeneracy of E$_{u}$ and E$_{g}$ modes in the high temperature phase or, even more interestingly, the possibility that the symmetry of the low temperature phase is lower than what has been inferred so far and accordingly, the modes corresponding to higher dimensional representations become non-degenerate.
\cite{SpartaEPJB01,ChoiPRB03} is reproduced.

\begin{figure}[t]
\centerline{
\epsfig{figure=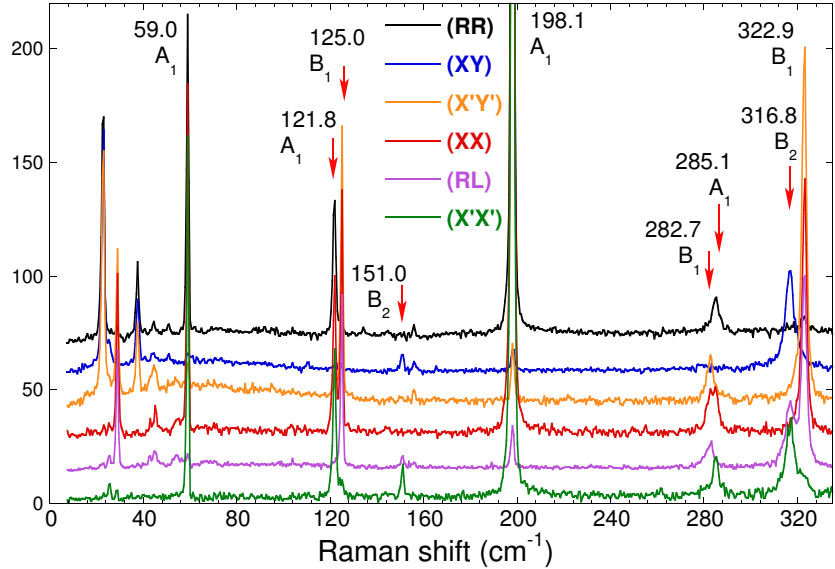,width=110mm}
}
\caption{
Raman data at T~$\approx$~3~K in six polarizations taken with the 1.92~eV laser excitation energy.
The energy and symmetry of the modes above 60~\cm-1 are shown in the figure.  
}
\label{f15}
\end{figure}

We show in Fig.~\ref{f15} our low temperature Raman data taken with in plane polarizations and using $\omega_{L}$~=~1.92~eV laser excitation energy.
In the $D_{2d}$ group the $(RR)$, $(RL)$, $(aa)$, $(ab)$, $(a'a')$ and $(a'b')$ probe $A_{1} + A_{2}$, $B_{1} + B_{2}$, $A_{1} + B_{1}$, $A_{2} + B_{2}$, $A_{1} + B_{2}$ and $A_{2} + B_{1}$ symmetries respectively.
The modes below 60~\cm-1 are not indexed since they are magnetic and will be discussed in the next section.
The modes are sharp and they are sitting at this low temperature on a flat background with almost vanishing intensity.
The 60~\cm-1 mode corresponds to the soft mode of the structural transition.
At 121.8~\cm-1 we observe in the A$_{1}$ channel the two phonon excitation seen also in $(cc)$ polarization, Fig.~\ref{f13}, and very close to it another sharp mode with $B_{1}$ symmetry.
We observe, similarly to the spectra shown in Fig.~\ref{f14}b, several pairs of modes having very similar energies.
For example doublet structures are observed around 284~\cm-1 where we see a pair of A$_{1}$ and B$_{1}$ excitations and also two modes having B$_{1}$ and B$_{2}$ symmetries are found around 320~\cm-1.

One way to explain this behavior is to follow up the suggestion in Ref.~\cite{ChoiPRB03} and assume that in the high temperature phase there are phonons which are odd and even with respect to inversion but very close in energy and to try to identify them by looking at similar atomic vibrations corresponding to 'u' and 'g' representations respectively.
The (A$_{1}$,B$_{1}$) group around 284~\cm-1 in Fig.~\ref{f15} would correspond in this scenario either to a group of (A$_{1g}$,A$_{1u}$) or (B$_{1g}$,B$_{1u}$) in the high temperature phase since A$_{1u}$ representation becomes B$_{1}$ and B$_{1u}$ representation becomes A$_{1}$ at low temperatures.
Similar reasoning would suggest for the (B$_{1}$,B$_{2}$) group around 320~\cm-1 that it originates either from a pair of (B$_{1g}$,A$_{2u}$) or (B$_{2g}$,A$_{1u}$) modes above 395~K.
We performed a symmetry analysis of the $k = 0$ atomic vibrations.
The conclusion is that this approach does not provide an a priori reason for the quasi-degeneracy and one has to perform a quantitative normal mode energy calculation by using appropriate inter-atomic elastic constants.
There is a simple way to see why this is true.
Looking at the character table of the $D_{4h}$ group it can be noticed that the even modes are symmetric with respect to the mirror symmetry in the Cu(BO$_{3}$) planes while the odd modes are antisymmetric with respect to this symmetry operation.
This means that the 'u' phonons in the high temperature phase correspond to vibrations of the atoms along the $c$-axis while the 'g' modes consist of in-plane movements.
Due to the different oscillation pattern it is hard to explain the closeness of phononic energies at this qualitative level.
We note that one can easily find vibrations corresponding to different group representations which involve similar oscillations at the 'molecular' level (for instance groups of atoms forming the Cu-O spin dimer structure or pairs O2 atoms between nearest neighbor dimers) and having a certain inter-molecular phase pattern, but they do not correspond to the experimentally observed symmetries.
We suggest that, remaining within the conclusions of X-ray studies which have not found evidence for additional crystallographic changes at low temperatures, good candidates for understanding this intriguing behavior are provided by the inter-dimer BO$_{3}$ molecular complexes whose rotations as a whole around the $a'$ and $c$-axes or whose in and out of the plane translations may turn out to be similar in energies.

\section{Magnetic Properties of \scbo}

\subsection{Experimental and Theoretical Reviews}

\begin{figure}[t]
\centerline{
\epsfig{figure=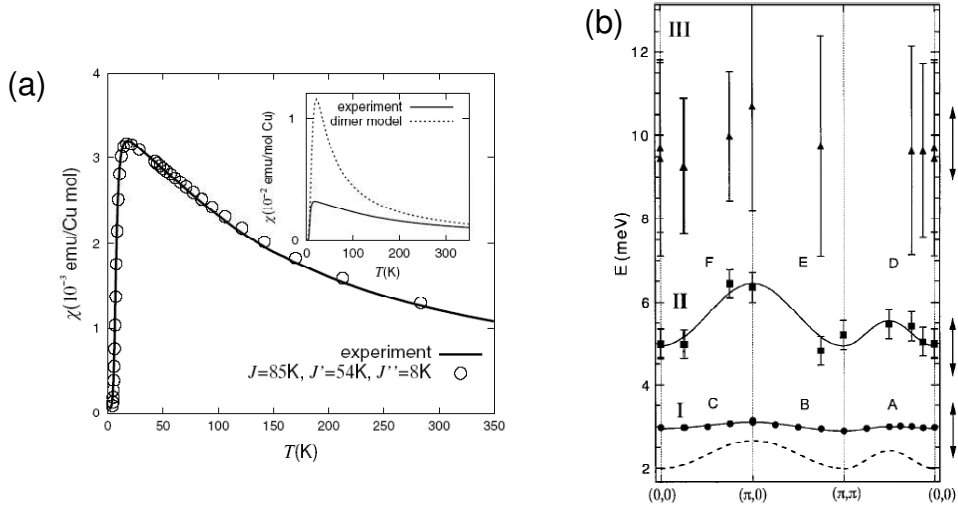,width=130mm}
}
\caption{
Temperature dependence of the magnetization (a) and inelastic neutron scattering results as a function of in plane wavevectors at low temperatures in \scbo (b) from Ref.~\cite{KageyamaPRL99}.
The solid line in (a) shows the result of a fit from numerical calculation using $x = J_{2}/J_{1}$~=~0.635, $J_{1}$~=~59~\cm-1 (7.32~meV) and also an interlayer coupling $J_{3}$~=~5.5~\cm-1. 
The interactions $J_{1}$ and $J_{2}$ alone provide a good description of the low temperature data, including specific heat, but a finite $J_{3}$ was necessary to explain the high temperature behavior of the magnetization where a Weiss temperature $\theta$~=~-92.5~K was obtained from the fit between 160 and 400~K~\cite{KageyamaPRL99}.
}
\label{f16}
\end{figure}

{\bf Experimental review --}
In this part we discuss data which relate to the most interesting properties and set the relevant energy scales of \scbo.
Magnetization and INS data provided for the first time evidence for the existence of a gapped phase in this compound~\cite{KageyamaPRL99}.
Fig.~\ref{f16}a shows that there is a drop in $\chi$ below about 15~K suggesting a gapped phase.
Another important thing pointed in the inset of this figure is the strong suppression of the magnetization peak around 20~K compared to the prediction of a simple dimer model which points towards the importance of other magnetic interactions. 
Neutron scattering, Fig.~\ref{f16}, probed the excitations out of this phase and confirmed the existence of a gap of about 3~meV (24~\cm-1) and found additional excitations around 5 and 9~meV.
Notable is the flat dispersion (less than 6\%) of the gap branch seen at 3~meV as a function of in-plane wavevectors meaning that the lowest excitations are very local.
The dispersion of the higher energy branches is more pronounced suggesting more mobile excitations in the multi-triplet channels.
The inability of the dimer model to describe the experimental data suggests that inter-dimer interactions are important and points towards a very interesting physics of frustration in this system.

Besides INS and magnetization, ESR and IR data in magnetic fields confirmed the magnetic nature of these excitations.
\begin{figure}[b]
\centerline{
\epsfig{figure=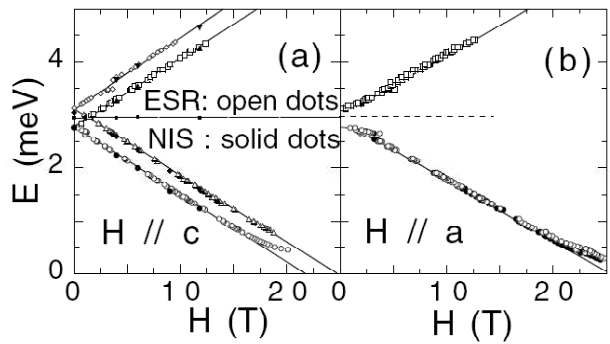,width=85mm}
}
\caption{
Neutron (Ref.~\cite{CepasPRL01}) and ESR (Ref.~\cite{NojiriJPSJ99}) data at low temperatures as a function of magnetic fields parallel (a) and perpendicular (b) to the $c$-axis. 
}
\label{f17}
\end{figure}
These experiments, along with high resolution INS data display a rich internal structure of the magnetic excitations and various selection rules for in and out of the plane applied magnetic fields, see Fig.~\ref{f17}.
In particular, the gap multiplet is shown to be made of 6 branches, which is expected since the unit cell contains 4 spins and therefore exciting a triplet on each of them while the other one is in a singlet state will give a total of 6 excitations.
Out of these 6 modes, 4 are seen to be symmetrically disposed around the gap value $\Delta$~=~24.2~\cm-1.
The observed splitting suggests that besides the superexchange interactions there are other low energy scale interactions which have to be taken in to account. 

{\bf Theoretical review --}
This section discuses the salient properties of the magnetic structure of \scbo and the excitations of the Hamiltonian~(\ref{e11}).
Among them, the ground state properties, the localization of the elementary triplets leading to the dispersionless magnon branches from Fig.~\ref{f16}b, the existence of strongly bound two-triplet states as well as the presence of additional spin orbit couplings generating the fine structure seen in Fig.~\ref{f17}.

In the approximation given by Eq.~(\ref{e11}) the wavefunction given by the direct product of singlet dimers is always an eigenstate of the system and it can also be shown that it is the ground state for a continuous set of parameters $x =J_{2}/J_{1}$.
For the proof of the eigenstate one has to take into account only the second term in (\ref{e11}) since the dimer product is obviously an eigenstate of the AF coupled independent dimers.
The application of the $J_{2} ({\bf S}_{1} \cdot {\bf S}_{3} + {\bf S}_{2}  \cdot {\bf S}_{3})$ on the $|s_{12}> \otimes |s_{34}>$ vanishes because the operator $({\bf S}_{1} + {\bf S}_{2})$ applied to the singlet state $|s_{12}>$ vanishes.
This is essentially due to the different parity of the singlet and triplet states with respect to the $1 \leftrightarrow 2$ exchange and the fact that on the orthogonal dimer lattice the Hamiltonian conserves the parity.
This remains true even if the interlayer coupling is considered~\cite{MiyaharaJPCM03}.

For $x \leq 0.5$ one can show that the singlet dimer product, $|\psi>$, is indeed the ground state.
The Hamiltonian (\ref{e11}) can be written as $\hat{H} = \sum_{i}^{N_{t}} \hat{h}_{i}$ where $N_{t}$ is the number of triangles of the type formed by the spins 1, 2 and 3 in Fig.~\ref{f19} and $\hat{h}_{i} = (J_{1}/2) ({\bf S}_{1} \cdot {\bf S}_{2}) + J_{2} ({\bf S}_{1} \cdot {\bf S}_{3} + {\bf S}_{2} \cdot {\bf S}_{3})$.
The ground state of each $\hat{h}_{i}$ is $e_{g}^{i} = -3/8 J_{1}$ if $x \leq 0.5$.
Accordingly, denoting the true ground state of $\hat{H}$ by $|\phi>$, one has $E_{g} = \  <\phi|\hat{H}|\phi> \ = \ \sum_{i}^{N_{t}} <\phi|\hat{h}_{i}|\phi> \ \geq -3/8 J_{1} N_{t}$ because $|\phi>$ is a variational function for $\hat{h}_{i}$.
Considered as a variational wavefunction for $\hat{H}$ and taking into account that the action of the second term in Eq.~(\ref{e11}) on $|\psi>$ is identically zero, one obtains immediately that $<\psi|\hat{H}|\psi> = -3/4 \ J_{1} N_{d} = -3/8 \ J_{1} N_{t}$ because the number of dimers, $N_{d}$, is half of that of the triangles.
On account of the variational principle, the true ground state energy satisfies $E_{g} \leq -3/8 J_{1} N_{t}$.
From the two inequalities one obtains that $E_{g} = -3/8 J_{1} N_{t}$ and so the product of singlet dimers is indeed the ground state.
The underlying reason for these beautiful properties is the fact that the magnetic lattice of \scbo is a realization of a 2D Shastry-Sutherland model which up to now has been discussed only at a theoretical level.

Turning to the question of the flat dispersion of the excitations in the one-triplet sector one can consider in Fig.~\ref{f19} the case of two nearest neighbor orthogonal dimers having an excited triplet state on the vertical bond and a singlet on the horizontal one.
The propagation of the triplet on the horizontal bond by the NNN term in Eq.~(\ref{e11}) is only possible if a triplet is left behind on the vertical dimer.
This is due to the reflection symmetry of singlet and triplet excitations with respect to a mirror plane parallel to the horizontal bond, Fig.~\ref{f19}a and the fact that the Hamiltonian of the system must be a fully symmetric operator.
If a triplet is left behind, then the hopping of one triplet to a neighboring dimer is very restricted and is possible only by virtually forming closed paths of triplets, the smallest of these paths involving three adjacent dimers.
As a result, one triplet hopping appears only in the 6$^{th}$ order of perturbation theory showing that these excitations are very localized in real space and explaining the flat $k$ dispersion seen in Fig.~\ref{f16}.
The motion of two triplets is different however.
It has been shown~\cite{KnetterPRL00} that in this case correlated hopping processes can occur and it was found that two-particle hopping appears in 2$^{nd}$ order perturbation theory.
\begin{figure}[t]
\centerline{
\epsfig{figure=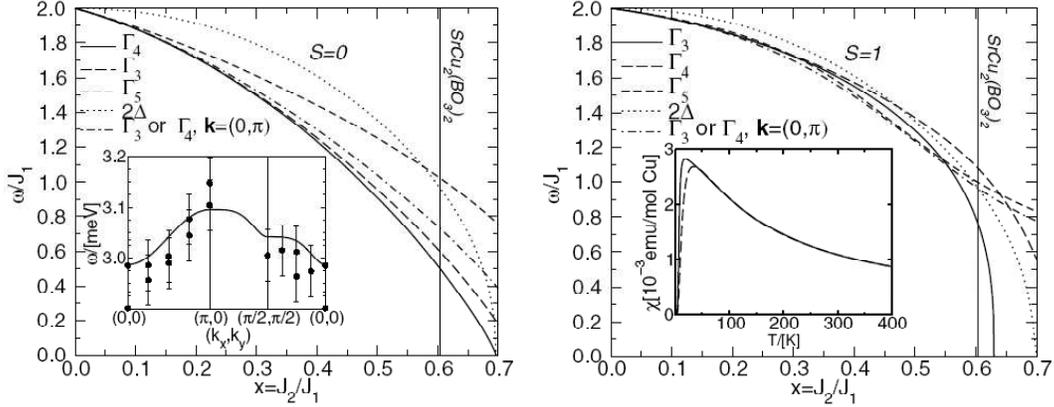,width=140mm}
}
\caption{
Predictions for the two-triplet bound states in the singlet (left) and triplet (right) of a perturbative analysis of the Shastry-Sutherland Hamiltonian of Eq.~(\ref{e11}) from Ref.~\cite{KnetterPRL00}.
$\Gamma_{1}$, $\Gamma_{2}$, $\Gamma_{3}$, $\Gamma_{4}$ and $\Gamma_{5}$ correspond to $A_{1}$, $B_{1}$, $B_{2}$, $A_{2}$ and $E$ in the notation of Table~\ref{t11}. 
}
\label{f18}
\end{figure}
This explains the larger dispersion seen by INS for the excitations around 5~meV. 

We saw that the products of dimer singlets is the ground state of $\hat{H}$ at least for $x \leq 0.5$.
What is the value of $x$ above which this is no longer true?
That there should be a finite value is obvious from the fact that in the limit $x \rightarrow \infty$ the lattice is topologically equivalent to a 2D square lattice which has long range order at T~=~0~K and a spin-wave approximation is more appropriate.
Theoretical work~\cite{MiyaharaJPCM03,KnetterPRL00,MiyaharaJPSJS00,MiyaharaPRL99} shows that below a ratio $x = J_{2} / J_{1} \approx 0.7$ the ground state remains the same and the system has a finite spin gap, $\Delta$ to the lowest excited $S = 1$ state.
The spin gap is equal to $J_{1}$ for $x = 0$ but with increasing this ratio $\Delta$ gets renormalized down due to many body effects, see Fig.~\ref{f18}.
At high values of $x$ the system has long range order, other possible intervening states separated by QCP's have been proposed to exist around $x = 0.7$.
The gap renormalization as a function of $x$ is shown in Fig.~\ref{f18}.
As shown here and also in the parameters used to fit magnetization data in Fig.~\ref{f16}, \scbo is closed to a QCP having a gap renormalized significantly, to a bit more than 50\% of the 'bare' value given by~$J_{1} \approx 85$~K (59~\cm-1 or 7.3 meV), see Ref.~\cite{MiyaharaJPCM03}.

Another characteristic of this magnetic lattice seen in Fig.~\ref{f18} is the existence of bound states in the two-triple sectors, see Fig.~\ref{f18}.
These are states which have an energy below the inset of the two-magnon continuum starting at $2 \Delta$.
Many of these have been predicted along with their symmetries.
The vertical line in Fig.~\ref{f18} showing the position of \scbo in the phase diagram was inferred from the experimentally found values of the spin gap $\Delta = 24$~\cm-1 (3~meV) and the observation by Raman scattering of a strong and sharp resonance (attributed to collective $S = 0$ two-triplet bound state) around 30~\cm-1~\cite{LemmensPRL00}.
We will discuss such excitations in the next section devoted to the analysis of low temperature Raman data in \scbo.

Before that, another observation in connection to the experimental results shown in Fig.~\ref{f17}:
there it is seen that the 6 branches of the gap multiplet around 24~\cm-1 are split in three pairs of doublets.
C\'{e}pas and collaborators proposed that this is due to the existence of inter-dimer DM interactions which have a direction parallel to the $c$-axis, see Fig.~\ref{f19}.
The proposed Hamiltonian (for the spins in the unit cell) reads:
\begin{equation}
\hat{h}_{c} = \vec{d}_{c}^{\ 13} ({\bf S}_{1} \times {\bf S}_{3}) + \vec{d}_{c}^{\ 14} ({\bf S}_{1} \times {\bf S}_{4}) + \vec{d}_{c}^{\ 23} ({\bf S}_{2} \times {\bf S}_{3}) + \vec{d}_{c}^{\ 24} ({\bf S}_{2} \times {\bf S}_{4}) 
\label{e12}
\end{equation}
The DM vectors satisfy $\vec{d}_{c}^{\ 13} = \vec{d}_{c}^{\ 24} = - \vec{d}_{c}^{\ 14} = - \vec{d}_{c}^{\ 23}$ due to the crystal symmetry, in particular due to the existence of mirror planes orthogonal to the dimers as shown in Fig.~\ref{f19} and the existence of $C_{2}$ rotation axes parallel to the $c$-axis and passing through the middle of the dimers.
It was found that this interaction reproduces the behavior in magnetic fields perpendicular and parallel to the $(ab)$ plane, which is shown by solid lines in Fig.~\ref{f17}.
\begin{figure}[t]
\centerline{
\epsfig{figure=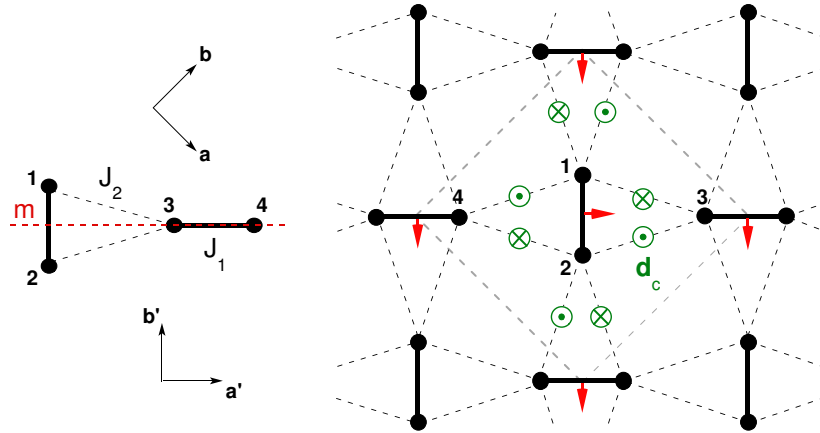,width=110mm}
}
\caption{
Left: Cartoon with two nearest neighbor dimers discussed in the text related to the exact ground state and hopping of triplet excitations.
$m$ denotes a mirror plane; the singlet and triplet states on the $1-2$ dimer will have different symmetries with respect to this reflection operator.
Right: The unit cell of the magnetic lattice from Fig.~\ref{f12} with the inter-dimer Dzyaloshinskii-Moriya term suggested in Ref~\cite{CepasPRL01} which is parallel to the $c$-axis.
The arrows perpendicular to each dimer correspond to our proposed antisymmetric intra-dimer interaction leading to singlet-triplet mixing. 
}
\label{f19}
\end{figure}
The energy of the upper and lower pairs is given by $\Delta \ \pm \ d_{c}$ and they remain degenerate, see also Fig~\ref{f111}c where we show symmetry analysis results.
Importantly, the inter-dimer DM interactions parallel to the $c$-axis are allowed both above and below the structural phase transition at 395~K.
Other DM terms, for instance the red arrows in Fig.~\ref{f19}, the  are not allowed above T$_{c}$ because the Cu(BO$_{3}$) plane is a mirror symmetry element.

\subsection{Magnetic Raman Scattering Results in \scbo}

Here are some experimental details related to our data discussed in the following. 
The spectra were taken from the $(ab)$ single crystal surface in a backscattering geometry.
We used an incident power of about 0.6~mW focussed on a 100~$\mu$ diameter spot.
The crystallographic axes orientation was determined by X-ray diffraction.
The data in magnetic fields, taken at a sample temperature of about  3~K, were acquired having the continuous flow cryostat inserted in the horizontal bore of a superconducting magnet.
We used the $\omega_{L} = $~1.92 and 2.6~eV excitation energies of a  Kr$^{+}$ laser and a triple-grating spectrometer for the analysis of the scattered light.
The data were corrected for the spectral response of the spectrometer and detector.
Polarization configurations are denoted by $({\bf e}_{in} {\bf e}_{out})$ where these two vectors are along the polarization direction of the incoming and outgoing photons.
Circular polarizations are denoted by $(RR)$ and $(RL)$ where ${\bf e}_{in,out} = (\hat{a} \pm i\hat{b})/\sqrt{2}$.

\begin{figure}[t]
\centerline{
\epsfig{figure=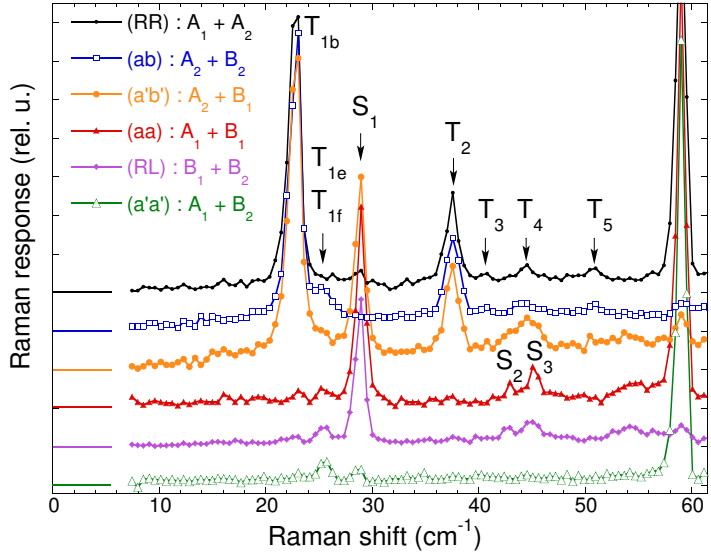,width=100mm}
}
\caption{
Zero field Raman data in \scbo taken with $\omega_{L} = 1.92$~eV excitation energy at T~=~3~K in six polarizations.
The legend shows the tetragonal symmetries probed in each scattering geometry. 
}
\label{f110}
\end{figure}

In Fig.~\ref{f110} we show six low temperature Raman spectra in zero applied field and using a laser frequency $\omega_{L} = 1.92$~eV.
The symmetries probed by each polarization in the tetragonal group are shown in the legend.
The mode seen at 60~\cm-1 is the soft mode of the structural transition, also seen in Fig.~\ref{f13}, which belongs as expected to the fully symmetric representation.
Three strong features are seen in the spectra at 23, 29 and 38~\cm-1 and they are denoted by T$_{1b}$, S$_{1}$ and T$_{2}$.
These excitations belong to the A$_{2}$, B$_{1}$ and A$_{2}$ representations respectively.
Besides these three modes, we observe several other weaker excitations.
Among them we see a set of three A$_{2}$ symmetric modes denoted by T$_{3}$, T$_{4}$ and T$_{5}$.
One can also observe the presence of other very weak feature at 25.5~\cm-1 in all polarization configurations except $(RR)$.
This suggest that at this frequency there are two quasi-degenerate excitations and that they belong to the B$_{1}$ and B$_{2}$ representations which would justify their observations in five out of six scattering geometries.
The symmetry analysis discussed later confirms indeed the above assumption.
The modes are denoted by T$_{1e}$ and T$_{1f}$.
A summary of the zero field energies and their experimentally observed symmetries can be found in Table~\ref{t12}.
The energy at which we observe these modes, the comparison with the INS, ESR and IR data, part of which are shown in Figs.~\ref{f16} and \ref{f17} as well as our data in magnetic fields confirm the magnetic origin of these modes and the predominant $S = 1$ character of the 'T' modes.
The T$_{1b}$, T$_{1e}$ and T$_{1f}$ excitations modes seem thus to belong to the spin gap multiplet while T$_{2}$, T$_{3}$, T$_{4}$ and T$_{5}$ would correspond to multi-particle $S = 1$ channels.

\begin{figure}[t]
\centerline{
\epsfig{figure=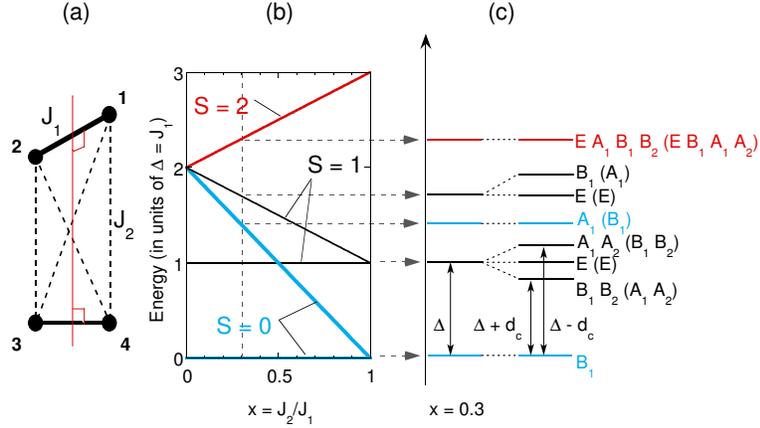,width=100mm}
}
\caption{
(a) The cluster used for the symmetry analysis of the $k = 0$ modes generated by the four spins in the unit cell of Fig.~\ref{f19}.
(b) The eigenvalues of the Hamiltonian corresponding to the 4-spin cluster in (a) taking into account the $J_{1}$ and $J_{2}$ terms as a function of the ratio $x = J_{2}/J_{1}$.
(c) Results of the symmetry analysis for this cluster in the $D_{2d}$ group.
The energy levels correspond to a value $x = 0.3$.
On the right we show the energy splittings and the absolute and relative to the ground state (in parenthesis) symmetries of the 16 magnetic modes when a finite inter-dimer DM interaction $d_{c}$ is present.
A finite intra-dimer DM term $d_{ab}$ will further split the modes which belong to the one-dimensional representations, see Fig.~\ref{f19}.
}
\label{f111}
\end{figure}

The zone center elementary excitations are generated by the spins within the magnetic unit cell.
If the picture of real space localized elementary triplets is true, then one expects that the analysis of the 4-spin cluster forming the unit cell is able to predict correctly the experimentally observed 
symmetries of these excitations.
Accordingly we calculated the energies and symmetries of the excitations generated by the magnetic cluster shown in Fig.~\ref{f111}a and the Hamiltonian of Eq.~(\ref{e11}).
The total number of states, 16, consist of two singlet, $S = 0$, states, three triplet, $S = 1$, states and one quintuplet, $S = 2$ state.
The energies are shown in Fig.~\ref{f111}a as a function of $x = J_{2}/J_{1}$ and one can observe their linear variation with the this ratio.
The points at $x = 0$ and $x = 1$ can be easily understood.
The first one corresponds to two independent AF coupled by $J_{1}$ on the bonds $1 \leftrightarrow 2$ and $3 \leftrightarrow 4$ so the energies will be just the sum of those corresponding to singlet and/or triplet states sitting on each dimer separately.
For the second point one can see that $[\hat{H},{\bf S}_{tot}] = 0$ so the energies will be, up to a constant factor, equal to the total spin eigenvalues $S(S + 1)$ (in units of $\hbar^{2}$).
Starting with four $S = 1/2$ spins one can have for the total spin the values $S = 0, 1$ and $2$.

The cluster in Fig.~\ref{f111}a belongs to the $D_{2d}$ group which is, as it should be, the point group associated with the space symmetry of the crystal.
We performed the group symmetry analysis of the eigenstates by decomposing the representation obtained starting with the canonical spin basis (the direct product of $|\uparrow>$ and $|\downarrow>$ spinors representing each Cu atom) into the irreducible representations of the $D_{2d}$ point group.
Each symmetry element of the $D_{2d}$ group involved the action on spinor state in the origin of the coordinate system as well as the corresponding permutation of spin indices.
Using projection operators we block diagonalized the Hamiltonian and numerical analysis (including the DM terms shown in Fig.~\ref{f19} and discussed in more detail later) allowed us to identify the energy and the predominant spin character of each eigenstate.

Indeed, as it is shown in the right-most column of the Table~\ref{t12} and Fig.~\ref{f111}c, the symmetries of the observed one-triplet excitations correspond to the results of group theory analysis.
The latter predicts for the six elementary triplets the A$_{1}$, A$_{2}$, B$_{1}$, B$_{2}$ and E symmetries, see Fig.~\ref{f111}.
The $E$ modes, T$_{1c}$ and T$_{1d}$, cannot be observed in zero field when the light propagates parallel to the $c$-axis because they are accessible only in $(ca)$ or $(cb)$ polarizations.
The fully symmetric T$_{1a}$ mode which, within the spin model including $J_{1}$, $J_{2}$ and the out-of-plane inter-dimer DM interaction proposed in Ref.~\cite{CepasPRL01}, should be degenerate with the strong T$_{1b}$ (A$_{2}$ symmetric) excitation at 22.8~\cm-1 is also not observed.
This is most probably due to a much weaker coupling to light in this symmetry channel.
The analysis shown in Fig.~\ref{f111} shows that the pair of T$_{1e}$ and T$_{1f}$ modes at 25.6~\cm-1 are to be seen in the B$_{1}$ and B$_{2}$ symmetry channels which explains the weak feature seen in Fig.~\ref{f110} in all polarizations except the $(RR)$ scattering geometry. 
The observation of the T$_{1e}$ (T$_{1f}$) modes with B$_{1}$ (B$_{2}$) symmetries 2.8~\cm-1 above the A$_{2}$ symmetric mode T$_{1e}$ allows the determination of the magnitude of the inter-dimer interaction $d_{c}$ (see Fig.~\ref{f111}) and also of its absolute sign because a sign change will interchange the position of these modes around the gap $\Delta$.
Finite intra-dimer Dzyaloshinskii-Moriya interactions (red arrows in Fig.~\ref{f19}) will split slightly the energies of the states in Fig.~\ref{f111} which belong to the one dimensional representations.
These effects will be discussed in more detail later where we analyze quantitatively the behavior of the spin excitations in magnetic fields applied parallel and perpendicular to the $c$-axis. 

We turn now to the discussion of two-triplet states.
Besides the singlet ground state and six one-triplet states, the symmetry analysis of the 4-spin cluster shown in Fig.~\ref{f111} predicts the following:
one $S = 0$ two-triplet bound state to be observed in the B$_{1g}$ channel, three branches with A$_{1}$ and double degenerate E symmetries which belong to another bound $S = 1$ excitation and five branches of a quintuplet ($S = 2$) state to be accessed in the A$_{1}$, A$_{2}$, B$_{1}$ and E channels.
\begin{table}[b]
\caption[]
{
Collective spin excitations in zero field: notation, the predominant 
spin character, $S_{tot}$ and $z$ projection $S_{z}$, energy and 
transition symmetries as observed experimentally and predicted from 
the 4-spin cluster in Fig.~1 corresponding to $k = 0$ excitations. 
}
\vspace{0.2cm}
\begin{center}
\begin{tabular}{c|c|c|ccc}
Mode & S$_{tot}$ (S$_{z}$) & Energy & \multicolumn{3}{c}{Symmetry} \\
  & & & Experiment & & Group Theory \\
\hline
T$_{1a}$ & $1$ ($\pm 1$) & 22.8 & - & \vline & A$_{1g}$ \\
T$_{1b}$ & $1$ ($\pm 1$) & 22.8 & A$_{2g}$ & \vline & A$_{2g}$ \\
T$_{1c}$ & $1$ ($0$) & 24.2 & - & \vline & E \\
T$_{1d}$ & $1$ ($0$) & 24.2 & - & \vline & E \\
T$_{1e}$ & $1$ ($\pm 1$) & 25.6 & B$_{1g}$ & \vline & B$_{1g}$ \\
T$_{1f}$ & $1$ ($\pm 1$) & 25.6 & B$_{2g}$ & \vline & B$_{2g}$ \\
\hline
S$_{1}$ & 0 ($0$) & 28.9 & B$_{1g}$ & \vline & - \\
T$_{2}$ & 1 & 37.5 & A$_{2g}$ & \vline & - \\
T$_{3}$ & 1 & 40.8 & A$_{2g}$ & \vline & - \\
S$_{2}$ & 0 & 43.0 & B$_{1g}$ & \vline & - \\
T$_{4}$ & 1 & 44.5 & A$_{2g}$ & \vline & - \\
S$_{3}$ & 0 & 45.2 & B$_{1g}$ & \vline & - \\
T$_{5}$ & 1 & 50.9 & A$_{2g}$ & \vline & - \\
\end{tabular}
\end{center}
\label{t12}
\end{table}
Their symmetries, along with the energies for the particular value $x = 0.3$ chosen as an example, are shown in Fig.~\ref{f111}c.
We observe that due to symmetry reasons none of the A$_{2}$ symmetric modes from T$_{2}$ to T$_{5}$ having energies higher than 30~\cm-1 qualify for an interpretation as triplet bound states generated 
within the 4-spin cluster.
This is consistent with the fact that larger cluster sizes are necessary in order to capture the more delocalized nature of these excitations which implies that they have contributions from the different parts of the Brillouin zone.
The fact that the existence of the strong A$_{2}$ symmetric bound triplet state at an energy 1.55~$\cdot \Delta$~=~37.5~\cm-1 has not been predicted by high order perturbative analysis~\cite{KnetterPRL00} suggest that other spin interactions have to be taken into account in order to explain the excitation 
spectrum.
Apparently, symmetry considerations would allow the 28.9~\cm-1 feature denoted by S$_{1}$ in Fig.~\ref{f110} to be interpreted as the singlet bound state of two triplets within a unit cell.
As we show in the following section, this 28.9~\cm-1 mode does not shift in external fields, which is compatible with a collective singlet excitation as discussed in~\cite{LemmensPRL00}, but suggests that its internal structure is not the one derived from the 4-spin cluster.

\begin{figure}[t]
\centerline{
\epsfig{figure=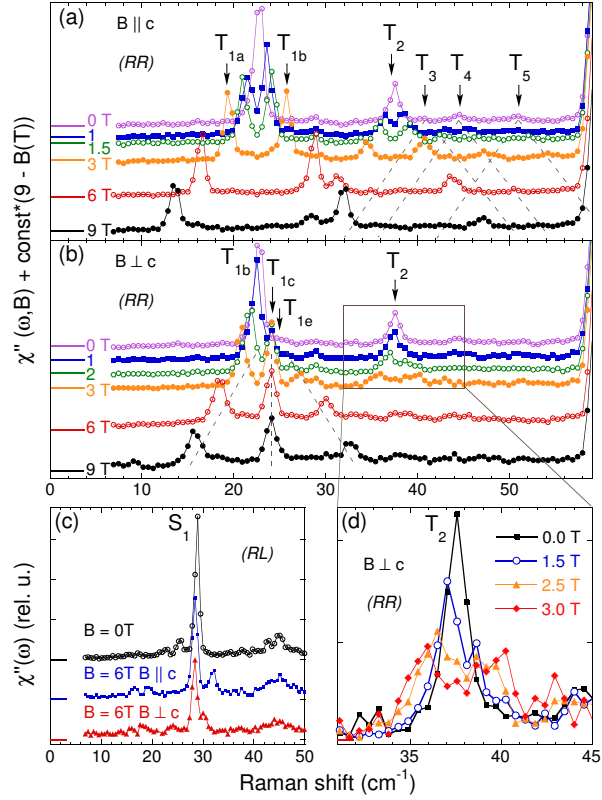,width=85mm}
}
\caption{
Magnetic field dependences of the magnetic excitations at T~=~3~K 
using the $\omega_{L}$~=~1.92~eV excitation energy in the following 
geometries:
(a) $(RR)$ $\vec{B} \parallel \hat{c}$.
(b) $(RR)$ $\vec{B} \perp \hat{c}$.
In (c) the $(RL)$ polarized data is shown for 0 and 6~T magnetic 
fields for both $\vec{B} \parallel \hat{c}$ and $\vec{B} \perp 
\hat{c}$.
In (a) and (b) the vertical shift is proportional to the magnetic 
field difference with respect to the 9~T spectrum and the dashed 
lines are guides for the eye.  
}
\label{f112}
\end{figure}

In Fig.~\ref{f112}, using the same mode notations, we show the influence of an external magnetic field applied parallel and perpendicular to the $c$-axis on the low temperature Raman spectra from Fig.~\ref{f110}.
The relevant aspects are the following.
In panel (a) we observe the splitting of the T$_{1a}$ and T$_{1b}$ modes in magnetic fields B~$\parallel c$, the B~=~1~T showing that the A$_{2}$ mode (T$_{1b}$) present in zero field disperses upwards 
with increasing the magnitude of the field.
Dashed lines in this figure mark the dispersion of the much weaker modes T$_{3}$, T$_{4}$ and T$_{5}$.
In Fig.~\ref{f112}b one of the E modes becomes Raman active due to symmetry lowering for $\vec{B} \perp \hat{c}$ configuration and we observe three dispersing branches of the gap multiplet.
Fig.~\ref{f112}c shows that the B$_{1}$ symmetric excitation at 28.9~\cm-1 does not change its energy with field, only a very small negative shift of the order of 0.5~\cm-1 from 0 to 6~T is seen because of the crossing with the upward dispersing gap branches seen in $(RR)$ polarization.
Panel (d), which is a zoomed in region of Fig.~\ref{f112}b, shows that several modes become Raman active in finite fields $\vec{B} \perp \hat{c}$ around 38~\cm-1 where the  T$_{2}$ excitation lies.
The internal structure of this higher energy multiplet is composed of modes dispersing up, down or independent of magnetic field.
Remarkable is their similar selection rules and dynamics in magnetic fields of the collective modes around 38~\cm-1  and 24~\cm-1.
The emergence in finite fields of several strong modes in the spin gap region precludes the observation of the weak T$_{1e}$ and T$_{1f}$ modes from Fig.~\ref{f110}.

\begin{figure}[h]
\centerline{
\epsfig{figure=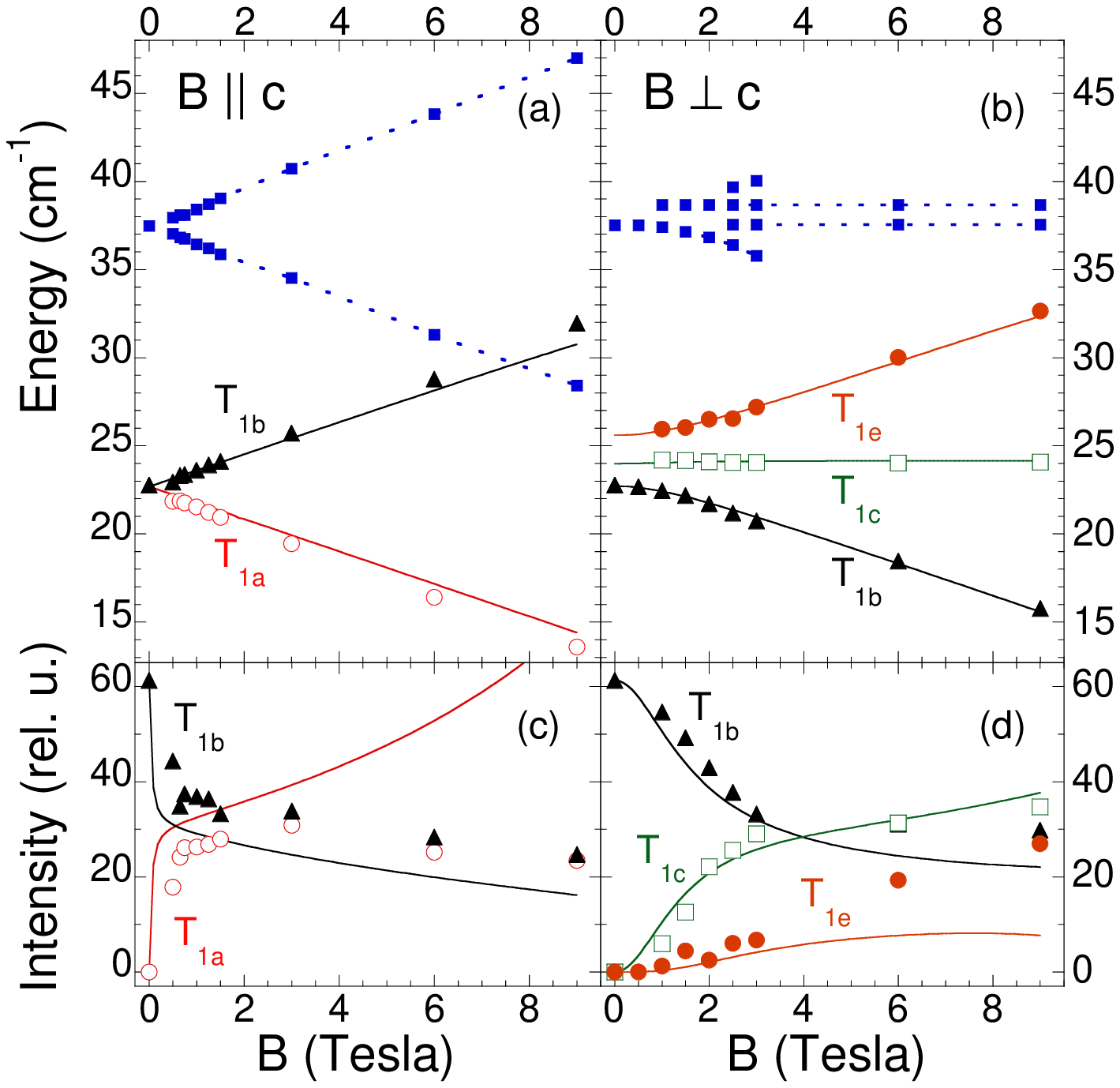,width=80mm}
}
\caption{
Energies (panels a and b) and intensities (panels c and d) of the spin excitations for $\vec{B} \parallel \hat{c}$ (right) and $\vec{B} \perp \hat{c}$ (left) from Fig.~3a-b.
Symbols represent experimental points, solid lines are the results of 4-spin cluster diagonalization as described in the text, dashed lines are guides for the eye. 
}
\label{f113}
\end{figure}

Fig.~\ref{f113} summarizes the magnetic field dependencies of the energies and spectral weights of the most intense Raman excitations.
The symbols in Fig.~\ref{f113}a-d correspond to experimental data, dashed lines are guides for the eye while the solid lines are results of a numerical diagonalization of a 4-spin cluster using the same set of parameters.
Taking into account that the 4-spin cluster neglects many-body gap renormalization effects (see Fig.~\ref{f18}) leading to a singlet-triplet energy independent of $x = J_{2}/J_{1}$ as well as the fact that when using periodic boundary conditions there is an effective doubling of the $J_{2}$ and inter-dimer DM interactions from Eq.~(\ref{e12}), we chose the following values: $J_{1}$~=~$\Delta$~=~24.2~\cm-1 which is the value of the spin gap, see the Table~\ref{t12}; 
$x = 0.556$ from the ratio of the energies of the sub-gap mode at 21.5~\cm-1~\cite{NojiriJPSJ03,ToomasPrivate} with respect to the gap $\Delta$ (see Fig.~\ref{f111}); an inter-dimer DM term parallel to the $c$-axis, $d_{c}$~=~1.4~\cm-1, which produces the splitting of the T$_{1a,b}$ and T$_{1e,f}$ branches from 24.2~\cm-1 (our value is consistent to the one proposed in the literature~\cite{CepasPRL01});
finally, from the magnetic field value around which the intensities crossing in Fig.~\ref{f113} takes place, the value chosen for the intra-dimer interaction was $d_{ab}$~=~2.66~\cm-1.

The term containing the intra-dimer DM interaction is proposed by us in order to explain the coupling and selection rules with magnetic fields.
Symmetry considerations impose in the $I\bar{4}2m$ group for the directions of the in plane DM vectors the ones depicted in Fig.~\ref{f19} with the Hamiltonian reading:
\begin{equation}
\hat{h}_{ab} = \vec{d}_{ab}^{\ 12} (\vec{S}_{1} \times \vec{S}_{2}) + \vec{d}_{ab}^{\ 34} (\vec{S}_{3} \times \vec{S}_{4})
\label{e13}
\end{equation}
The interaction terms responsible for the coupling to the external photon field and which were used for the calculation of intensities in Fig.~\ref{f113}c-d are discussed in the next part.
We also plotted (filled squares and dashed lines) the field dependence of other higher energy modes observed in Fig.~\ref{f112}.

We remark an overall qualitative agreement for this choice of parameters which suggests that the intra-dimer interaction has to be taken into account in the spin Hamiltonian.
The agreement is quantitative as regards the energies and the intensity variations for the $\vec{B} \perp \hat{c}$ case.
The term described in Eq.~(\ref{e13}) plays a crucial role in obtaining a finite coupling to the excited $S = 1$ triplets which is not realized by the $d_{c}$ terms in Eq.~(\ref{e12}).
Regarding the apparent degeneracy of the T$_{1a}$ and T$_{1b}$  as well as that of the T$_{1e}$ and T$_{1f}$ modes although they belong to different representations, we note that these two groups of excitations are split by $\hat{h}_{ab}$ but the splitting is very small, of the order of $d_{ab}^{\ 2} / \Delta \approx$~0.25~\cm-1.
The largest discrepancy between the experimental data and the calculation is seen in Fig.~\ref{f113}.
One aspect in this regard is that the value $d_{ab}$ had to be chosen 
greater than that of $d_{c}$.
This is intriguing because the $d_{c}$ term is allowed by symmetry both above and below the structural phase transition at 395~K~\cite{SpartaEPJB01} whereas the existence of a finite intra-dimer 
DM interaction is allowed only below 395~K when the mirror symmetry of the $(ab)$ plane is just slightly broken.
Additional terms may be responsible for this disagreement, possible candidates being in-plane components of the inter-dimer DM interaction, which should also be allowed below the structural phase transition.

We discuss now the issue related to the existence of a magnetic mode \emph{below} the spin gap value~\cite{NojiriJPSJ03,ToomasPrivate}.
In order to reproduce the upward dispersion with fields $\vec{B} \parallel \hat{c}$ of the T$_{1b}$ mode we had to choose a value of $x$ which is greater than 0.5, otherwise this excitation would have 
displayed a downward dispersion.
From Fig.~\ref{f111} we observe that a ratio of $x = J_{2}/J_{1} \geq$~0.5 implies that the position of the bound singlet state is below $\Delta$.
We suggest that this state is responsible for the observations of the 21.5~\cm-1 mode in Refs.~\cite{NojiriJPSJ03,ToomasPrivate}.
The presence of this excitation will also influence the specific heat measurements and, in conjunction with the finite intra-dimer interaction $d_{ab}$, also the low temperature magnetization data 
whose quantitative understanding has not been achieved yet~\cite{MiyaharaJPCM03}.
The existence of this magnetic mode is at odds with theoretical predictions~\cite{KnetterPRL00,MiyaharaJPSJS00,MiyaharaPRL99}.
However, the experimental finding of the set of A$_{2}$ symmetric modes (T$_{2}$ to T$_{5}$ in Fig.~2, all of them below the two-magnon continuum starting at 2$\Delta \approx$~48~\cm-1 and also not 
predicted by theory) already shows that the understanding of the spin dynamics in higher particle sectors is not complete.

\begin{figure}[h]
\centerline{
\epsfig{figure=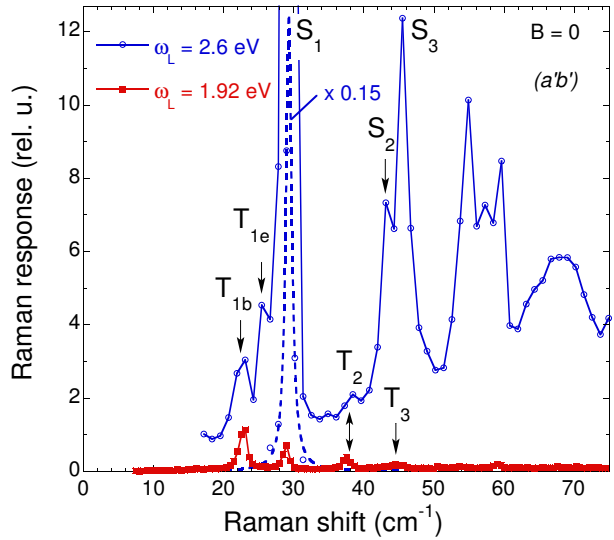,width=80mm}
}
\caption{
T~=~3~K Raman data in $(a'b')$ polarization for $\omega_{L}$~=~1.92 (filled squares) and 2.6~eV (empty circles) excitation energies.
The data points corresponding to the resonantly enhanced singlet bound state at 29~\cm-1 are multiplied by 0.15 and the corresponding line represents a Lorentzian fit. 
}
\label{f114}
\end{figure}

Fig.~\ref{f114} shows two low temperature Raman spectra taken in $(a'b')$ polarization with two incoming laser frequencies, $\omega_{L}$~=~1.92 and 2.6~eV.
The point we make is that we observe two types of behaviors.
Firstly we notice that the intensities of the group of A$_{2}$ symmetric modes, T$_{1b}$ and T$_{2}$ is about the same for the two photon energies used.
On the other hand, the modes corresponding to the group formed by the T$_{e}$, S$_{1}$, S$_{2}$ and S$_{3}$ resonances are more than two orders of magnitude stronger for $\omega_{L} = $~2.6~eV than for 
$\omega_{L} = $~1.92~eV.
Other modes in the 50 to 70~\cm-1 energy range also become visible in the $\omega_{L}$~=~2.6~eV spectrum.

Our data prove that the coupling to these two groups of excitations takes place by two distinct light scattering mechanisms.
The resonance of the B$_{1}$ symmetric T$_{e}$ magnetic mode, enhanced for $\omega_{L}$~=~2.6~eV, is similar to the one corresponding to S$_{1}$, S$_{2}$ and S$_{3}$ excitations as well as to the behavior of the new modes seen around 55, 59 and 68~\cm-1.
On the account of this similarity alone, one cannot identify the latter as magnetic bound states as well.
However, the lack of energy shift in magnetic fields and the results of perturbational analysis regarding energy scales and symmetries~\cite{KnetterPRL00} are not contradicting their interpretation as Raman active collective $S = 0$ magnetic bound states~\cite{LemmensPRL00}.

We discuss below the nature of the two light coupling mechanisms to magnetic excitations.
As for the set of A$_{2}$ symmetric modes we propose that the coupling takes place $via$ the spin-orbit coupling which can be written in an effective form as $({\bf e}_{in} \times {\bf e}_{out}) \vec{S}_{tot}^{\ z}$~\cite{FleuryPR68}.
This interaction Hamiltonian probes excitations with A$_{2}$ symmetry and the calculated magnetic field dependence in Fig.~\ref{f113} is also in agreement with the experimental results.
The coupling to the T$_{1e}$ and T$_{1f}$ modes from Fig.~\ref{f110} can be understood if we invoke the usual effective spin interaction corresponding to the photon induced spin exchange process $\sum_{<i,j>} ({\bf e}_{in} \cdot {\bf r}_{ij}) ({\bf e}_{out} \cdot {\bf r}_{ij}) \vec{S}_{i} \cdot \vec{S}_{j}$.
Here the sum runs over pairs of lattice sites, $\vec{S}_{i}$ and $\vec{S}_{j}$ are the exchanged spins on sites $i$ and $j$ respectively, while {\bf r}$_{ij}$ is the vector connecting these sites~\cite{FleuryPR68}.
Writing down the explicit expression of this interaction for several polarizations in the 4-spin cluster approximation (Figs.~\ref{f19} and \ref{f111}) one indeed gets finite coupling in B$_{1}$ and B$_{2}$ channels for the triplet T$_{1e}$ and T$_{1f}$ states.
This explains the presence of the 25.6~\cm-1 modes in all polarizations except $(RR)$.

The difference in the coupling strengths seen in Fig.~\ref{f114} is thus understandable because these two coupling mechanisms need not be simultaneously in resonance with the same high energy excited 
electronic states.
The photon induced spin exchange Hamiltonian has been usually invoked in order to explain Raman active $S = 0$ two-magnon type excitations in various magnetic systems~\cite{FleuryPR68}.
\scbo is an example where this Hamiltonian, in the presence of singlet-triplet mixing DM interactions, can be used to account for coupling to $S = 1$ states.
A remaining question is why do we not see the bound singlet mode with an energy below $\Delta$ for any of the two excitations used?
In principle the photon induced spin exchange, resonant in this case at higher photon energies, could provide coupling to this excitation also, as to the 28.9~\cm-1 mode denoted by S$_{1}$.
One possible explanation to be explored in more detail from a theoretical point of view is that the Raman form factors for exciting a pair of magnons both at $k = 0$ is vanishing as opposed to the case of a pair of zone boundary modes.
For instance this is the case when the Raman vertex is calculated for the 2D square lattice within the spin-wave approximation and using the Fleury-Loudon interaction Hamiltonian.
Consequently, the 21.5 and the 28.9~\cm-1 excitations could be both attributed to $S = 0$ bound states originating from different parts of the reciprocal space and having substantially different binding energies.  

\subsection{Summary}
We showed in this chapter low temperature Raman data on phononic  and magnetic excitations in \scbo.
Regarding the former, in the 0 to 350~\cm-1 range we find several pairs of quasi-degenerate modes which have different symmetries.
Group theoretical analysis suggests that the existence of these modes is related to quite different atomic vibrational pattern, i.e. in-plane and $c$-axis motions and as a result a quantitative investigation would be very interesting.
Collective magnetic excitations were studied in terms of symmetry, resonance and coupling 
mechanisms in zero and applied magnetic fields.
The analysis of the 4-spin cluster shown in Figs.~\ref{f19} and \ref{f111} allows us to understand the group symmetries of the zero field Brillouin zone center spin gap branches around 24~\cm-1 confirming the picture of local elementary one-triplet modes.
By considering an additional intra-dimer DM interaction we are also able to understand the observed selection rules and intensity variations of the spin gap branches in external magnetic fields applied parallel or perpendicular to the dimer planes.
These selection rules also require that the energy of the $S = 0$ two-triplet bound state made out of magnons confined within a unit cell is \emph{below} $\Delta$ (in the 4-spin cluster this is equivalent to $x \geq 0.5$ in Fig.~\ref{f110}) suggesting a very high binding energy for this two particle excitation.
The 4-spin cluster analysis fails to account for the two-triplet excitations which shows that they have contributions from the different parts of the reciprocal space.
The existence of a set of four modes below the onset of two-triplet continuum, at 37.5, 40.8, 44.5 and 50.9~\cm-1, in the A$_{2}$ symmetry channel, shows that further theoretical analysis is required in order to understand the nature of these composite excitations.
Finally, we identified two effective magnetic light scattering Hamiltonians responsible for the coupling to the magnetic modes which allowed us to explain their resonance behavior.

{\bf Acknowledgments --}
We acknowledge discussions and collaborations with B. S. Dennis, M. V. Klein and T. R\~{o}\~{o}m. 
The crystals were provided by H. Kageyama.


\end{document}